\documentclass[runningheads]{llncs}

\usepackage[T1]{fontenc}

\usepackage{graphicx}

\usepackage[british]{babel}

\usepackage[numbers]{natbib}
\usepackage{amsmath}
\usepackage{amssymb}

\usepackage{booktabs}
\usepackage{multirow}

\usepackage{mathpartir}

\usepackage{stmaryrd}

\usepackage{xcolor}

\usepackage{tikz}
\usepackage{pgfplots}

\usepackage[inline]{enumitem}

\usepackage{array}
\usepackage{multirow}

\usepackage{listings}
\definecolor{isarblue}{HTML}{006699}
\definecolor{isargreen}{HTML}{009966}
\lstdefinelanguage{isabelle}{%
    keywords=[1]{type_synonym,datatype,fun,function,abbreviation,definition,proof,lemma,theorem,corollary,inductive},
    keywordstyle=[1]\bfseries\color{isarblue},
    keywords=[2]{where,assumes,shows,and,fixes},
    keywordstyle=[2]\bfseries\color{isargreen},
    keywords=[3]{if,then,else,case,of,SOME,let,in,O},
    keywordstyle=[3]\color{isarblue},
}
\makeatletter
\lst@AddToHook{OnEmptyLine}{\vspace{-0.4\baselineskip}}
\makeatother

\usepackage{etoolbox}
\makeatletter
\patchcmd{\lsthk@TextStyle}{\let\lst@DefEsc\@empty}{}{}{\errmessage{failed to patch}}
\makeatother

\usepackage{calc}

\usepackage{todonotes}

\usepackage[hidelinks]{hyperref}
\usepackage[capitalise]{cleveref}

\newcommand{\MLSS}{\textbf{MLSS}}

\newlength{\trianglewidth}
\newcommand{\lefttrianglebar}{\mathrel{\includegraphics{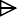}}}
\newcommand{\lefttriangle}{\mathrel{\includegraphics{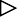}}}

\newcommand{\lexpands}[2]{#1 $\lefttriangle$ #2}
\newcommand{\fexpands}[2]{#1 $\lefttrianglebar$ #2}

\newcommand{\expandssSym}{\lefttriangle^*}
\newcommand{\expandss}[2]{#1 $\expandssSym$ #2}

\newcommand{\unionS}{\sqcup_\text{s}}
\newcommand{\interS}{\sqcap_\text{s}}
\newcommand{\diffS}{-_\text{s}}
\newcommand{\inS}{\in_\text{s}}
\newcommand{\notinS}{\notin_\text{s}}
\newcommand{\eqS}{=_\text{s}}
\newcommand{\neqS}{\neq_\text{s}}

\newcommand{\Ist}{I$_\text{st}$}

\newcommand{\hfmem}{\boldsymbol{\in}}

\newcommand{\fmAndSymbol}{\boldsymbol{\land}}
\newcommand{\fmAnd}[2]{#1 $\fmAndSymbol$ #2}
\newcommand{\fmOr}[2]{#1 $\boldsymbol{\lor}$ #2}
\newcommand{\fmNegSymbol}{\boldsymbol{\neg}}
\newcommand{\fmNeg}[1]{$\fmNegSymbol\:$#1}
\newcommand{\fmAtom}{\textbf{A}}

\begin{document}


\setcounter{tocdepth}{1}

\title{Towards a Verified Tableau Prover for a Quantifier-Free Fragment of Set Theory}
\titlerunning{Towards a Verified Tableau Prover for MLSS}
\author{Lukas Stevens\orcidID{0000-0003-0222-6858}}
\institute{Technical University of Munich, Boltzmannstr.\ 3, 85748 Garching, Germany
  \email{lukas.stevens@in.tum.de}}

\maketitle

\begin{abstract}
  Using Isabelle/HOL, we verify the state-of-the-art decision procedure for multi-level syllogistic with singleton (\MLSS{} for short), which is a quantifier-free fragment of set theory.
  We formalise its syntax and semantics as well as a sound and complete tableau calculus for it.
  We also provide an executable specification of a decision procedure that exhaustively applies the rules of the calculus and prove its termination.
  Furthermore, we extend the calculus with a lightweight type system that paves the way for an integration of the procedure into Isabelle/HOL.
\end{abstract}

\section{Introduction}
In Isabelle/HOL, there are specialised procedures for dealing with e.g.\ natural numbers, linear arithmetic, and metric spaces.
Some of these procedures have been verified in Isabelle/HOL, such as a procedure for Presburger arithmetic~\cite{presburger} that was later extended to mixed real-integer arithmetic~\cite{arithmetic}.
This procedure, though, uses reflection to work on goals in Isabelle/HOL, which, during execution, either sacrifices speed by going through the simplifier or requires trusting the code generator.
More recently, \citeauthor{orders}~\cite{orders} presented a verified decision procedure for orders that produces certificates. 
This approach offers efficient execution by using generated code as well as soundness because the certificates are replayed through Isabelle's inference kernel.

This paper focuses on another ubiquitous structure in mathematics, namely sets.
To the best of our knowledge, we present the first formally verified decision procedure for (a fragment of) set theory.
In particular, we consider a quantifier-free fragment which \citeauthor{new_fast_tableau}~\cite{new_fast_tableau} call multi-level syllogistic with singleton (\MLSS{}).
The fragment includes the usual set operations of union, intersection, difference, membership, equality and, in addition, it allows the construction of singleton sets.

Since \MLSS{} admits a tableau calculus, generating certificates will be straightforward.
Like with the aforementioned order solver, this paves the way for an integration of the decision procedure into Isabelle, adding to its growing body of verified decision procedures.

\subsection{Contributions}
We present a formalisation in Isabelle/HOL of a tableau calculus for \MLSS{} due to \citeauthor{new_fast_tableau}~\cite{new_fast_tableau}\cite[Chapter 14]{set_theory}.
We prove soundness and completeness of the calculus and give an abstract specification of a decision procedure that exhaustively applies the rules of the calculus.
To obtain total correctness of the procedure, we prove its termination.
Additionally, we naively refine the abstract to an executable specification from which we can generate code.
The formalisation initially follows the paper but offers a more thorough account of some important details:
\begin{itemize}
  \item We deliver the omitted proof of Lemma~2 in the paper~\cite{new_fast_tableau}, a key building block for the completeness proof of the calculus.
  \item The formal proof of completeness reveals that the calculus lacks a rule for eliminating double negation. 
  \item We derive an explicit upper bound for the number of formulas in a tableau branch.
\end{itemize}

In the context of Isabelle/HOL, there is one crucial aspect that requires us to modify the calculus in the paper:
the calculus works under the assumption that every variable is a set;
however, this is not the case in Isabelle/HOL, e.g.\ consider the expression \lstinline!n $\in$ A! where \lstinline!n! is a natural number.
We call these variables urelements.
To deal with them, we extend the calculus with a lightweight type system and a verified inference algorithm that identifies the urelements.

The modification of the calculus required non-trivial changes to the completeness proof.
Here, the formalisation was instrumental because Isabelle immediately revealed which proofs had been broken.
This illustrates the usefulness of ITPs for developing logic calculi: they allow us to confidently make modifications without compromising correctness. 

All in all, the formalisation amounts to over 6000 lines of theory.
It is part of the \emph{Archive of Formal Proofs} (AFP)~\cite{formalisation}.
The entry provides an overview theory \href{https://www.isa-afp.org/theories/mlss_decision_proc/#MLSS_Proc_All}{\texttt{MLSS\_Proc\_All.thy}} that highlights the (mostly syntactic) differences between paper and formalisation and references the constants and theorems that are introduced in this paper. 

\subsection{Related Work}
Since the literature on decidable fragments of set theory is vast, we only focus on \MLSS{} here.
\citeauthor{mlss_first}~\cite{mlss_first} were the first to show the decidability of the fragment.
Subsequent work~\cite{mlss_np} found the decision problem to be \textbf{NP}-complete.
To obtain a practical decision procedure, \citeauthor{mlss_first_tableau}~\cite{mlss_first_tableau} proposed a tableau calculus, which was later improved by \citeauthor{tableau_quantifier_free}~\cite{tableau_quantifier_free}.
Both of these procedures construct a model during execution that guides the proof search.
\citeauthor{tableau_quantifier_free} also cover an extension of the calculus with uninterpreted functions, which \citeauthor{mlss_quantification}~\cite{mlss_quantification} later revisited while avoiding the construction of a model during execution.
In this paper, we consider a version of the latter procedure due to \citeauthor{new_fast_tableau}~\cite{new_fast_tableau} that is specialised to \MLSS{} and where the branching rules of the calculus are set up to guarantee the mutual exclusivity of the branches.
Later extensions of the calculus added certain interpreted functions, such as monotone functions~\cite{mlss_monotone_functions} and the inverse of a function~\cite{mlss_cartesian_map}.
The latter extension notably includes the Cartesian product.
Those extensions, though, did not improve upon the tableau calculus for \MLSS{}.

There is a large body of work at the intersection of ITPs and tableau methods, but to keep with this paper's theme we only consider formalisations of correctness here.
For first-order logic, there are abstract completeness proofs using the \emph{Beth-Hintikka style} of possibly infinite derivation trees~\cite{completeness_coinductive} as well as the \emph{Henkin style} of maximally consistent sets~\cite{synthetic_completeness_afp}.
Both are abstract enough to be instantiated with a wide range of concrete calculi.
A more concrete formalisation~\cite{sequent_first_order} verifies a sequent calculus for first-order logic whose completeness proof is via a translation to semantic tableau.

Beyond completeness, we target decidability, which is more attainable for propositional logic.  
There is a verified tableau calculus for the modal logic S5~\cite{lean_modal_logic} in Lean and one for hybrid logic~\cite{hybrid_logic} in Isabelle/HOL.
Both of these do not prove termination but there is a formalisation of a tableau calculus for the temporal logic CTL in Coq~\cite{ctl_tableau} that does.

\subsection{Notation}
Isabelle/HOL~\cite{isabelle} conforms to everyday mathematical notation for the most part.
We establish notation and in particular some essential data types together with their primitive operations that are specific to Isabelle/HOL.

We write \lstinline!t :: 'a! to specify that the term \lstinline!t! has the type \lstinline!'a! and \lstinline!'a $\Rightarrow$ 'b! for the space of total functions from type \lstinline!'a! to type \lstinline!'b!.

Sets with elements of type \lstinline!'a! have the type \lstinline!'a set!.
The cardinality of a set \lstinline!A! is denoted by \lstinline!|A|! and the image of \lstinline!A! under \lstinline!f! by \lstinline!f ` A!.

We use \lstinline!'a list! to describe the type of lists, which are constructed using the empty list \lstinline![]! constructor or the infix cons constructor \lstinline!#!, and are appended with the infix operator \lstinline!@!.
The function \lstinline!set! converts a list into a set.

We remark that $\longleftrightarrow$ is equivalent to \lstinline!$=$! on the type of Booleans \lstinline!bool! and \lstinline!$\equiv$! is definitional equality of the meta-logic of Isabelle/HOL, which is called Isabelle/Pure.
Meta-implication is denoted by \lstinline!$\Longrightarrow$! and a chain of implications
\lstinline!A$_\text{1}$ $\Longrightarrow$ $\cdots$ $\Longrightarrow$ A$_\text{k}$ $\Longrightarrow$ C!
can be abbreviated by 
\lstinline!$\llbracket$ A$_\text{1}$;$\,\ldots\,$;A$_\text{k}$ $\rrbracket$ $\Longrightarrow$ C!.

\section{Syntax and Semantics of MLSS\label[section]{sec:semantics}}
\subsection{Syntax}
At the heart of \MLSS{}, we have the type of set terms, which is the disjoint union of the empty set and variables as well as the operations union, intersection, difference, and the singleton set represented by the constructor \lstinline!Single!.
We keep the type of variables abstract by making it a parameter of the set term data type.
The only restriction on the type of variables is that it needs to be infinite.
Isabelle/HOL's data type package automatically defines a function that gives us the set of variables in a set term, which we name \lstinline!vars!.
In what follows, we will overload the function \lstinline!vars! to also work on set atoms, formulas, and branches.
\begin{lstlisting}
datatype (vars: 'a) pset_term =
    $\emptyset$ | Var 'a | Single ('a pset_term)
  | 'a pset_term $\unionS$ 'a pset_term
  | 'a pset_term $\interS$ 'a pset_term
  | 'a pset_term $\diffS$ 'a pset_term
\end{lstlisting}
We can combine two set terms to form a set atom by using the membership or the equality operator.
\begin{lstlisting}
datatype (vars: 'a) pset_atom =
    'a pset_term $\inS$ 'a pset_term
  | 'a pset_term $\eqS$ 'a pset_term
\end{lstlisting}
With the above operators we can also represent the subset operator \lstinline!$\subseteqS$! and enumerate finite sets: \lstinline!s $\subseteqS$ t! is equivalent to \lstinline!s $\unionS$ t $\eqS$ t! and a finite set of elements \lstinline!{t$_\text{1}$,$\ldots$,t$_\text{k}$}! can be expressed by \lstinline!Single t$_1$ $\unionS$ $\ldots$ $\unionS$ Single t$_k$!.

We use the propositional fragment of formulas due to \citeauthor{lqe}~\cite{lqe} with set atoms as propositional atoms to form the quantifier-free fragment \MLSS{} of set theory.
\begin{lstlisting}
datatype (atoms: 'a) fm =
    &\fmAtom& 'a
  | &\fmNeg{('a fm)}&
  | &\fmAnd{'a fm}{'a fm}&
  | &\fmOr{'a fm}{'a fm}&

type_synonym 'a pset_fm = 'a pset_atom fm
\end{lstlisting}
We will often drop the atom constructor \lstinline!&\fmAtom{}&! to reduce clutter.
Additionally, we use \lstinline!s $\notinS$ t! and \lstinline!s $\neqS$ t! to denote \lstinline!&\fmNeg{\fmAtom{} (s $\inS$ t)}&! and \lstinline!&\fmNeg{\fmAtom{} (s $\eqS$ t)}&!, respectively.

Similarly to \lstinline!vars!, we get the function \lstinline!atoms :: 'a fm $\Rightarrow$ 'a set! for free that retrieves all set atoms in a formula.
We combine these functions to extract all the variables occurring in a set formula.
\begin{lstlisting}
definition vars $\phi$ $\equiv$ $\bigcup$(vars ` atoms $\phi$)
\end{lstlisting}

Likewise, we fix the constant \lstinline!subterms :: 'b $\Rightarrow$ 'a pset_term set! that is polymorphic in its argument type \lstinline!'b!.
We overload this constant to return the set terms that are subterms of a set term, set atom, or formula, respectively.
Lastly, we introduce the function \lstinline!subfms :: 'a fm $\Rightarrow$ 'a fm set! that computes the subformulas of a formula.
The functions \lstinline!subterms! and \lstinline!subfms! are implemented in the expected way.

\subsection{Semantics}
The original paper~\cite{new_fast_tableau} bases the semantics of \MLSS{} on the von Neumann hierarchy of sets $\mathcal{V}$.
We instead use the hierarchy of \emph{hereditarily finite sets} (HF sets) which fulfil all the same axioms as $\mathcal{V}$ -- that is, the axioms of ZF -- except for the axiom of infinity.
In particular, the membership relation is well-founded.
The HF sets, as we will see, are sufficient to construct a model for any satisfiable \MLSS{} formula.
In contrast to $\mathcal{V}$, the HF sets are directly representable in Isabelle/HOL, and indeed, an AFP entry~\cite{hf_AFP} formalises them.
The entry defines a type \lstinline!hf! that comes with the following functionality:
\begin{itemize}
  \item The function \lstinline!HF :: hf set $\Rightarrow$ set! that converts a finite set of HF sets into an HF set.
\item The usual set operations such as equality ($=$), membership ($\hfmem$), union ($\sqcup$), intersection ($\sqcap$), and difference ($-$) are defined.
\item Finally, the empty set coincides with the ordinal $0$, so it is denoted by \lstinline!0 :: hf!.
\end{itemize}

Equipped with the above, we define the interpretation functions 
\begin{itemize}
  \item \lstinline!&\Ist& :: ('a $\Rightarrow$ hf) $\Rightarrow$ 'a pset_term $\Rightarrow$ hf! and 
  \item \lstinline!&\Isa& :: ('a $\Rightarrow$ hf) $\Rightarrow$ 'a pset_atom $\Rightarrow$ hf!
\end{itemize}
in the standard way, i.e.\ by mapping each syntactic construct to the corresponding operation on HF sets and interpreting variables with respect to a given valuation function \lstinline!M :: 'a $\Rightarrow$ hf!.
For the concrete definition we refer to the formalisation.

We write \lstinline!M $\models$ $\phi$! for the judgement that the formula $\phi$ holds under the valuation function \lstinline!M!.
The implementation of $\models$ coincides with the interpretation function of \citeauthor{lqe}~\cite{lqe}.
As usual, we call a formula \lstinline!$\phi$! \emph{satisfiable} if there exists a model \lstinline!M! with \lstinline!M $\models$ $\phi$!.
Otherwise, we say that \lstinline!$\phi$! is \emph{unsatisfiable}.

\section{A Tableau Calculus for MLSS}
We formalise the tableau calculus for \MLSS{} as described by \citeauthor{new_fast_tableau}~\cite{new_fast_tableau}.
Inspired by the formalisation of a tableau calculus for hybrid logic by \citeauthor{hybrid_logic_afp}~\cite{hybrid_logic_afp}, we use lists to represent the branches of the tableau tree.
Note that we add formulas to the front of the list during branch expansion, so \lstinline!last b! for a branch \lstinline!b! is always the formula we are trying to disprove with the tableau.
We sometimes call this formula the \emph{initial formula}.
\begin{lstlisting}
type_synonym 'a branch = 'a pset_fm list
\end{lstlisting}

\noindent We lift the functions \lstinline!vars! and \lstinline!subterms! to branches in the expected way.

In the standard tableau calculus for propositional logic as \citeauthor{tableau}~\cite{tableau} describes it, a branch is called \emph{closed} if it contains both the negation of a formula and the formula itself;
conversely, it is called \emph{open} if it is not closed.
For \MLSS{}, we extend the notion of closedness with three additional rules; the first two are straightforward while the last one states that a branch is closed when the branch contains a membership cycle
\lstinline!t$_\text{0}$ $\inS$ t$_\text{1}$, t$_\text{1}$ $\inS$ t$_\text{2}$, $\ldots$, t$_\text{k}$ $\inS$ t$_\text{0}$!.

\begin{lstlisting}
inductive bclosed :: 'a branch $\Rightarrow$ bool where
  $\llbracket$ $\phi$ $\in$ set b; &\fmNeg{$\phi$}& $\in$ set b $\rrbracket$ $\Longrightarrow$ bclosed b
| (t $\inS$ $\emptyset$) $\in$ set b $\Longrightarrow$ bclosed b
| (t $\neqS$ t) $\in$ set b $\Longrightarrow$ bclosed b
| $\llbracket$ member_cycle cs; set cs $\subseteq$ set b$\rrbracket$ $\Longrightarrow$ bclosed b

abbreviation bopen b $\equiv$ $\neg$ bclosed b
\end{lstlisting}
A tableau is called \emph{closed} if all of its branches are closed.

\subsection{Linear Expansion Rules}
\begin{table*}[t]
  \centering
  \lstset{
    escapeinside={@}{@}
  }
  \newcommand{\ton}{\text{t}_\text{1}}
  \newcommand{\ttw}{\text{t}_\text{2}}
  \caption{Linear expansion rules.\label[table]{tab:lexpands} All rules except the double negation rule coincide with the original paper~\cite{new_fast_tableau}. For brevity, we omit the rules for $\interS$ and $\diffS$.}
  \begin{tabular}{ccc}
    \toprule
    Propositional Rules & {\hskip 6pt} & Rules for \lstinline!$\unionS$! \\
    \cmidrule(l{1pt}){1-1}\cmidrule(r{1pt}){3-3}
    \begin{tabular}{p{0.18\textwidth}m{0.05\textwidth}p{0.175\textwidth}}
      \lstinline!@\fmAnd{p}{q}@! & $\Longrightarrow$ & \lstinline!p, q! \\
      \lstinline!@\fmNeg{(\fmOr{p}{q})}@! & $\Longrightarrow$ & \lstinline!@\fmNeg{p}@, @\fmNeg{q}@! \\
      \lstinline!@\fmOr{p}{q}, \fmNeg{p}@! & $\Longrightarrow$ & \lstinline!q! \\
      \lstinline!@\fmOr{p}{q}, \fmNeg{q}@! & $\Longrightarrow$ & \lstinline!p! \\
      \lstinline!@\fmNeg{(\fmAnd{p}{q})}@, p! & $\Longrightarrow$ & \lstinline!@\fmNeg{q}@! \\
      \lstinline!@\fmNeg{(\fmAnd{p}{q})}@, q! & $\Longrightarrow$ & \lstinline!@\fmNeg{p}@! \\
      \lstinline!@\fmNeg{(\fmNeg{p})}@! & $\Longrightarrow$ & \lstinline!p!\\[6ex]
    \end{tabular}
                        & &
    \begin{tabular}{p{0.22\textwidth}m{0.05\textwidth}p{0.22\textwidth}}
    \lstinline!s $\notinS$ $\ton$ $\unionS$ $\ttw$! & $\Longrightarrow$ & \lstinline!s $\notinS$ $\ton$, s $\notinS$ $\ttw$! \\
    \lstinline!s $\inS$ $\ton$! & $\Longrightarrow$ & \lstinline!s $\inS$ $\ton$ $\unionS$ $\ttw$! \\
    \lstinline!s $\inS$ $\ttw$! & $\Longrightarrow$ & \lstinline!s $\inS$ $\ton$ $\unionS$ $\ttw$! \\
    \lstinline!s $\inS$ $\ton$ $\unionS$ $\ttw$!, \newline \lstinline!s $\notinS$ $\ton$! & $\Longrightarrow$ & \lstinline!s $\inS$ $\ttw$! \\
    \lstinline!s $\inS$ $\ton$ $\unionS$ $\ttw$!, \newline \lstinline!s $\notinS$ $\ttw$! & $\Longrightarrow$ & \lstinline!s $\inS$ $\ton$! \\
    \lstinline!s $\notinS$ $\ton$, s $\notinS$ $\ttw$! & $\Longrightarrow$ & \lstinline!s $\notinS$ $\ton$ $\unionS$ $\ttw$! \\[0.9em]
    \end{tabular}
    \\

    Rules for \lstinline!$\interS$! && Rules for \lstinline!$\diffS$! \\
    \cmidrule(l{1pt}){1-1}\cmidrule(r{1pt}){3-3}
    \begin{tabular}{c}
      $\vdots$
    \end{tabular}
    & &
    \begin{tabular}{c}
      $\vdots$ \\
    \end{tabular}
    \\[0.9em]

    Rules for \lstinline!Single! && Rules for \lstinline!$\eqS$! \\
    \cmidrule(l{1pt}){1-1}\cmidrule(r{1pt}){3-3}
    \begin{tabular}{p{0.18\textwidth}m{0.05\textwidth}p{0.175\textwidth}}
      & $\Longrightarrow$ & \lstinline!s $\inS$ Single s! \\
      \lstinline!s $\inS$ Single t! & $\Longrightarrow$ & \lstinline!s $\eqS$ t! \\
      \lstinline!s $\notinS$ Single t! & $\Longrightarrow$ & \lstinline!s $\neqS$ t! \\
    \end{tabular}
    & &
    \begin{tabular}{p{0.22\textwidth}m{0.05\textwidth}p{0.22\textwidth}}
      \lstinline!$\ton$ $\eqS$ $\ttw$, l! & $\Longrightarrow$ & \lstinline!l{$\ttw$/$\ton$}! \\
      \lstinline!$\ton$ $\eqS$ $\ttw$, l! & $\Longrightarrow$ & \lstinline!l{$\ton$/$\ttw$}! \\
      \lstinline!s$_\text{1}$ $\inS$ t, s$_\text{2}$ $\notinS$ t! & $\Longrightarrow$ & \lstinline!s$_\text{1}$ $\neqS$ s$_\text{2}$! \\
    \end{tabular}
    \\
    \bottomrule
  \end{tabular}
\end{table*}

The calculus considers two kinds of branch expansion rules: \emph{linear} and \emph{branching} rules.
As the name suggests, branching rules lead to the creation of new branches in the tableau while linear rules only extend a branch \lstinline!b! with new formulas \lstinline!b' = [$\psi_1$,$\ldots$,$\psi_n$]!, which we denote by \lstinline!&\lexpands{b'}{b}&!.
\cref{tab:lexpands} shows the linear expansion rules.
Note that in the first two rules for \lstinline!$\eqS$!, \lstinline!l! is a literal occurring in the branch.
Furthermore, the term-for-term substitution \lstinline!l{s/t}! is restricted to the top-level set terms of \lstinline!l!, i.e.\ the set terms that occur directly under one of the atom constructors \lstinline!$\inS$! or \lstinline!$\eqS$!;
for example, given the literal
\begin{lstlisting}
  l = &\fmNeg{((s $\unionS$ u) $\diffS$ s $\eqS$ s $\unionS$ u)}&
\end{lstlisting}
we have
\begin{lstlisting}
  &\phantom{=}& (&\fmNeg{((s $\unionS$ u) $\diffS$ s $\eqS$ s $\unionS$ u)}&){t/s $\unionS$ u}
  = &\fmNeg{((s $\unionS$ u) $\diffS$ s $\eqS$ t)}&.
\end{lstlisting}
A more crucial restriction of the linear rules is that no new subterm may be created by their application;
for instance, the second rule for \lstinline!$\unionS$! is
\begin{lstlisting}
  s $\inS$ t$_\text{1}$ $\Longrightarrow$ s $\inS$ t$_\text{1}$ $\unionS$ t$_\text{2}$,
\end{lstlisting}
which formally represents 
\begin{lstlisting}
  (s $\inS$ t$_\text{1}$) $\in$ set b $\Longrightarrow$ &\lexpands{[s $\inS$ t$_\text{1}$ $\unionS$ t$_\text{2}$]}{b}&,
\end{lstlisting}
and may only be used under the condition
\lstinline[breaklines]!t$_\text{1}$ $\unionS$ t$_\text{2}$ $\in$ subterms (last b)!.
The purpose of this restriction is to prevent unbounded expansion of the branch.
In fact, we give an explicit upper bound for the number of formulas in a branch in \cref{sec:correct}.

Due to boundedness, repeated expansion with linear rules eventually results in a \emph{linearly saturated} branch, i.e.\ a branch where no application of linear rules would produce new formulas.
\begin{lstlisting}
definition lin_sat b $\equiv$ $\forall$b'. &\lexpands{b'}{b}& $\longrightarrow$ set b' $\subseteq$ set b
\end{lstlisting}

Finally, we remark that the original paper~\cite{new_fast_tableau} is missing the last propositional rule dealing with double negation.
This rule is required for completeness, though, considering that the branch
\lstinline![&\fmNeg{\fmNeg{\fmNeg{p}}}&, p, &\fmAnd{\fmNeg{\fmNeg{\fmNeg{p}}}}{p}&]!
is saturated --- neither linear nor branching rules apply --- and open, but there clearly is no model for the initial formula \lstinline!&\fmAnd{\fmNeg{\fmNeg{\fmNeg{p}}}}{p}&!.

\subsection{Branching Rules\label[section]{sec:branching}}
\begin{table*}[t]
  \lstset{
    escapeinside={@}{@}
  }
  \centering
  \newcommand{\ton}{\text{t}_\text{1}}
  \newcommand{\ttw}{\text{t}_\text{2}}
  \newcommand{\lpipe}{\rule[-0.4ex]{0.6pt}{2.3ex}}
  \caption{Branching expansion rules.\label[table]{tab:bexpands} We write \lstinline!$\phi$! for \lstinline!last b! here. All rules coincide with the original paper~\cite{new_fast_tableau} so we only show an illustrative subset.}
  \begin{tabular}{>{\centering\arraybackslash}p{0.34\textwidth} >{\centering\arraybackslash}p{0.32\textwidth} >{\centering\arraybackslash}p{0.3\textwidth}}
    \toprule
    Rule & Precondition & Subsumption condition \\
    \midrule
    \begin{tabular}{@{}>{\raggedleft\arraybackslash}p{0.11\textwidth}c>{\raggedright\arraybackslash}p{0.11\textwidth}@{}}
      \\
      \midrule
      \lstinline!p! & \lpipe & \lstinline!@\fmNeg{p}@! \\
    \end{tabular}
         & \lstinline!@\fmOr{p}{q}@ $\in$ set b! &
         \begin{tabular}{rl}
         & \lstinline!p $\in$ set b $\lor$! \\
         $\fmNegSymbol$ & \lstinline!p $\in$ set b! \\
         \end{tabular} \\

    \begin{tabular}{@{}>{\raggedleft\arraybackslash}p{0.11\textwidth}c>{\raggedright\arraybackslash}p{0.11\textwidth}@{}}
      \\
      \midrule
      \lstinline!s $\inS$ $\ton$! & \lpipe & \lstinline!s $\notinS$ $\ton$! \\
    \end{tabular}
         &
         \begin{tabular}{c}
           \lstinline!(s $\inS$ $\ton$ $\unionS$ $\ttw$) $\in$ set  b! \\
           \lstinline!$\ton$ $\unionS$ $\ttw$ $\in$ subterms $\phi$! \\
         \end{tabular}
         &
         \begin{tabular}{@{}rl@{}}
                  & \lstinline!(s $\inS$ $\ton$) $\in$ set b! \\
           $\lor$ & \lstinline!(s $\notinS$ $\ton$) $\in$ set b!
         \end{tabular}
         \\

      $\vdots$ & $\vdots$ & $\vdots$ \\

      \addlinespace[0.6em]

      \begin{tabular}{@{}>{\raggedleft\arraybackslash}p{0.165\textwidth}c>{\raggedright\arraybackslash}p{0.165\textwidth}@{}}
      \\
      \midrule
      \lstinline!Var x $\inS$ $\ton$! \lstinline!Var x $\notinS$ $\ttw$! & \multirow{2}{*}{\rule[-0.4ex]{0.6pt}{4.8ex}} & \lstinline!Var x $\notinS$ $\ton$! \lstinline!Var x $\inS$ $\ttw$! \\
    \end{tabular}
         & 
         \begin{tabular}{c}
           \lstinline!($\ton$ $\neqS$ $\ttw$) $\in$ set b! \\
           \lstinline!$\ton$ $\in$ subterms $\phi$! \\
           \lstinline!$\ttw$ $\in$ subterms $\phi$! \\
           \lstinline!x $\notin$ vars b! \\
         \end{tabular}
         &
         \begin{tabular}{r@{\hspace{0.5em}}l}

           \lstinline!$\exists$s.! & \lstinline!(s $\inS$ $\ton$) $\in$ set b! \\
               \lstinline!$\land$! & \lstinline!(s $\notinS$ $\ttw$) $\in$ set b! \\
           \multicolumn{2}{c}{$\lor$} \\
           \lstinline!$\exists$s.! & \lstinline!(s $\notinS$ $\ton$) $\in$ set b! \\
               \lstinline!$\land$! & \lstinline!(s $\inS$ $\ttw$) $\in$ set b! \\
         \end{tabular}
         \\
         \bottomrule
  \end{tabular}
\end{table*}
After running out of linear rules to apply, only the branching rules shown in \cref{tab:bexpands} remain.
A rule is applicable if its \emph{precondition} is met and, to prevent unnecessary branching, if it is not subsumed as indicated by the \emph{subsumption condition}.
These rules create multiple branches in the tableau, so we represent the different possibilities \lstinline!bs'! to expand a branch \lstinline!b! as a set and write \lstinline!&\fexpands{bs'}{b}&!. 
Accordingly, we get a new branch \lstinline!b' @ b! in the tableau for each \lstinline!b' $\in$ bs'!.

A linearly saturated branch where no further branching is possible is called a \emph{saturated} branch. 
\begin{lstlisting}
definition sat b $\equiv$ lin_sat b $\land$ ($\nexists$bs'. &\fexpands{bs'}{b}&)
\end{lstlisting}

Note that even branching rules are defined such that they never create new subterms, except for the last rule that adds a new variable to the branch.
These variables serve to manifest an inequality; hence, we call them \emph{witnesses}.
\begin{lstlisting}
definition wits b $\equiv$ vars b - vars (last b)
\end{lstlisting}

\section{A Decision Procedure for MLSS}
The mechanics of the decision procedure are typical for a procedure based on a tableau calculus:
it decides the satisfiability of a given formula \lstinline!$\phi$! by determining whether the formula has a closed tableau.
More specifically, it initialises the tableau with the singleton branch \lstinline![$\phi$]! and checks whether this branch can be expanded to a closed tableau.

We only discuss the abstract specification here and refer the reader to the formalisation for the executable specification.
The implementation uses a couple of features of Isabelle/HOL's function package:
instead of defining the function via pattern matching, we specify the equations of the function as conditional rewrite rules.
This requires us to prove that the assumptions of the equations are non-overlapping, which is done by automation.
The other concern is that Isabelle/HOL requires functions to be total, so a recursive function needs to terminate for it to be well-defined;
nevertheless, the termination proof is separated from the definition of the function for modularity.
The function package maintains the soundness of the definition by introducing a so-called domain predicate \lstinline!mlss_proc_branch_dom! which characterises the arguments for which the function terminates.
Each equation of the function is guarded by an assumption that the predicate holds for the argument.
In \cref{sec:correct}, we will show that the domain predicate holds for the context in which the function \lstinline!mlss_proc_branch! is called in.
Before we go into more detail on how the termination is proved, we discuss the definition of the function, as shown below.

\begin{lstlisting}
function mlss_proc_branch :: 'a branch $\Rightarrow$ bool where
  $\neg$ lin_sat b $\Longrightarrow$ mlss_proc_branch b =
  mlss_proc_branch ((SOME b'. &\lexpands{b'}{b}& $\land$
                              set b $\subset$ set (b' @ b)) @ b)
| $\llbracket$ lin_sat b; bclosed b $\rrbracket$ $\Longrightarrow$ mlss_proc_branch b = True
| $\llbracket$ $\neg$ sat b; bopen b; lin_sat b $\rrbracket$ $\Longrightarrow$ mlss_proc_branch b =
  ($\forall$b' $\in$ (SOME bs. &\fexpands{bs}{b}&). mlss_proc_branch (b' @ b))
| $\llbracket$ lin_sat b; sat b $\rrbracket$ $\Longrightarrow$ mlss_proc_branch b = bclosed b

definition mlss_proc :: 'a pset_fm $\Rightarrow$ bool where
  mlss_proc $\phi$ $\equiv$ mlss_proc_branch [$\phi$]
\end{lstlisting}

The purpose of the function is to determine whether we can expand a given branch to a closed tableau.
As stated before, we first use linear expansion rules in order to prevent premature branching;
to this end, we recursively expand the branch with linear rules until the branch is linearly saturated.
Note that we use Hilbert's $\varepsilon$-operator in the form of \lstinline!SOME!\footnote{In the formalisation, the function \lstinline!mlss_proc_branch! is actually parametrised by choice functions to allow for refinement.} to choose some rule that actually adds new formulas to the branch. 
As soon as the branch is linearly saturated, we terminate if the branch is closed as the second equation shows.
Otherwise, we choose an applicable branching rule and recursively check whether all newly created branches can be closed.
The final equation applies once no further branch expansion is possible, in which case we just test for closedness of the branch.

The procedure \lstinline!mlss_proc! then calls \lstinline!mlss_proc_branch! with a singleton branch \lstinline![$\phi$]! to determine the satisfiability of a given formula \lstinline!$\phi$!.

Thus, we use \lstinline!mlss_proc_branch! is only on branches that result from applying the expansion rules.
We call this kind of branch \emph{well-formed}.
In the definition below, the expression \lstinline!&\expandss{b'}{b}&! denotes that \lstinline!b'! is one of the branches that results from applying (potentially zero) expansion rules to \lstinline!b!.
\begin{lstlisting}
definition wf_branch b $\equiv$ $\exists$$\phi$. &\expandss{b}{[$\phi$]}&
\end{lstlisting}

We use this notion in \cref{sec:correct} to state an upper bound for the cardinality of well-formed branches.
The upper bound justifies the termination of the decision procedure. 
Before we come to that, though, we prove soundness and completeness in Section~\ref{sec:sound} and~\ref{sec:complete}, respectively.
In \cref{sec:correct}, we also show that both properties easily transfer to \lstinline!mlss_proc!, which, together with termination, establishes that it is a decision procedure.

\section{Completeness of the Calculus\label[section]{sec:complete}}
For completeness of the calculus, we need to show that every unsatisfiable formula has a closed tableau or, conversely, that the formula is satisfiable if there is a saturated and open branch in the tableau.
To facilitate inductive reasoning, we show a stronger statement by constructing a model \lstinline!M! such that \lstinline!M $\models$ $\phi$! for all \lstinline!$\phi$ $\in$ set b!.
At the core of the model, there is a \emph{realisation} function that maps set terms to sets of type \lstinline!hf!.
A subset of the witnesses, which we call \emph{pure} witnesses, receives special treatment from the realisation function for reasons that will become apparent in \cref{sec:pwits}.
The collection of set terms of a branch can thus be partitioned into two collections, as defined below.
\begin{lstlisting}
definition pwits :: 'a branch $\Rightarrow$ 'a set where
  pwits b $\equiv$ {c $\in$ wits b. $\forall$t $\in$ subterms (last b).
         $\phantom{\equiv}$     AT (Var c $\eqS$ t) $\notin$ set b $\land$ AT (t $\eqS$ Var c) $\notin$ set b} 

definition subterms' :: 'a branch $\Rightarrow$ 'a pset_term set where
  subterms' b $\equiv$ subterms (last b) $\cup$ Var ` (wits b - pwits b)
\end{lstlisting}

We aim to construct a syntactic model that we derive from the membership literals \lstinline!s $\inS$ t! in the branch.
To this end, we construct a graph whose vertices are the disjoint union of the sets above and there is an edge from \lstinline!s! to \lstinline!t! in the graph if, and only if, \lstinline!s $\inS$ t! is in \lstinline!b!.
Note that we use \citeauthor{graph_theory_afp}'s graph library~\cite{graph_theory_afp} which represents a graph as a record of vertices, arcs (directed edges), and two functions \lstinline!tail! and \lstinline!head! that map an arc to its source and target vertex, respectively.
\begin{lstlisting}
definition bgraph b $\equiv$ let vs = Var ` pwits b $\cup$ subterms' b
  in $\llparenthesis$ verts = vs, arcs = {(s, t). (s $\inS$ t) $\in$ set b},
     &$\phantom{\llparenthesis}$& tail = fst, head = snd $\rrparenthesis$
\end{lstlisting}

The realisation function is defined relative to this graph.
As mentioned before, the realisation function treats the pure witnesses differently than the rest of the set terms.
The function evaluates terms in the latter set in accordance to the structure of the graph, i.e.\ the realisation of a vertex is defined as the union of the realisations of the parent vertices. 
For the former set, we choose a function \lstinline!I! that assigns the pure witnesses pairwise distinct sets with cardinality greater than that of the vertices.
We can always choose such a function since we assume an infinite universe of variables.
Then, we return the singleton set \lstinline!HF {I x}!, which, together with the cardinality constraint, guarantees that realisations are distinct between pure witnesses themselves as well as between pure witnesses and set terms.  
The notation \lstinline!u $\rightarrow_\text{G}$ s! in the definition below indicates that there is an edge from \lstinline!u! to \lstinline!s! in the graph \lstinline!G!.
\begin{lstlisting}
abbreviation parents G s $\equiv$ {u. u $\rightarrow_\text{G}$ s}

function realise :: 'a pset_term $\Rightarrow$ V where
  x $\in$ Var ` pwits b $\Longrightarrow$ realise x = HF {I x}
| x $\in$ subterms' b
  $\Longrightarrow$ realise t = HF {realise ` parents (bgraph b) s}
| x $\notin$ verts G $\Longrightarrow$ realise x = 0
\end{lstlisting}

Again, we need to ensure that the assumptions of the equations are non-overlapping and that the function terminates.
The former is taken care of by automation, leaving us to prove termination.
The assumption that \lstinline!b! is open implies that there are no membership cycles, thus \lstinline!bgraph b! is acyclic.
Furthermore, the graph is finite by definition.
Thus, we can use the cardinality of the set of ancestors as a measure that decreases in each recursive call.

Before we prove that the realisation function constitutes a model in \cref{sec:real}, we will first explain the significance of the pure witnesses.

\subsection{Characterisation of the Pure Witnesses\label[section]{sec:pwits}}
Recall that the pure witnesses of a branch \lstinline!b! are those witnesses that are not related to other subterms in \lstinline!last b! by equality.
In the context of a well-formed branch, we can strengthen this characterisation to any set term and, in addition, we also get that there is no membership literal where a pure witness is on the right-hand side.
Intuitively speaking, the realisation of a pure witness does not depend on the realisation of any other set term.
\begin{lstlisting}[label={lst:lemma_2}]
lemma lemma_2:
  assumes wf_branch b and c $\in$ pwits b
  shows (Var c $\eqS$ t) $\notin$ set b and (t $\eqS$ Var c) $\notin$ set b
    and (t $\inS$ Var c) $\notin$ set b
\end{lstlisting}
\newcommand{\reflemmatwo}{\hyperref[lst:lemma_2]{\lstinline!lemma_2!}}
So why are pure witnesses treated differently?
According to the definition of \lstinline!realise!, it would evaluate the pure witnesses would to the empty set \lstinline!0 :: hf!, were they not treated separately. 
To see that this is a problem, consider the branch
\lstinline!b = [Var s $\neqS$ Var t, Var t $\neqS$ Var u]!
which expands to several open and saturated branches, one of which is
\begin{lstlisting}
  [Var x $\neqS$ Var y, Var x $\inS$ Var s, Var x $\notinS$ Var t,
  &\phantom{[Var x $\neqS$ Var y,}& Var y $\inS$ Var t, Var y $\notinS$ Var u] @ b
\end{lstlisting}
for some fresh \lstinline!x! and \lstinline!y!.
Assigning both \lstinline!Var x! and \lstinline!Var y! a value of \lstinline!0! would contradict the literal \lstinline!Var x $\neqS$ Var y!.
To prevent this, we assign the pure witnesses pairwise different values.

The proof of \reflemmatwo{} is more technical than interesting so we refer the reader to the formalisation.

\subsection{Realisation of an Open Branch\label[section]{sec:real}}
Remember that for completeness, we need to show that the realisation function for an open and saturated branch \lstinline!b! actually constitutes a model for all formulas in the branch.
We start by verifying that the realisation function models all literals in the branch; more formally, the following propositions hold:
\begin{enumerate}[label={(\arabic*)}]
  \item\label{it:at_mem} We have \lstinline!realise s $\hfmem$ realise t! if it holds that \lstinline!s $\inS$ t! is in \lstinline!b!. 
  \item\label{it:at_eq} We have \lstinline!realise s $=$ realise t! if \lstinline!s $\eqS$ t! is in \lstinline!b!.
  \item\label{it:af_eq} We have \lstinline!realise s $\neq$ realise t! if \lstinline!s $\neqS$ t! is in \lstinline!b!.
  \item\label{it:af_mem} We have \lstinline!realise s $\hfnotmem$ realise t! if it holds that \lstinline!s $\notinS$ t! is in \lstinline!b!. 
\end{enumerate}
To illustrate the usefulness of \reflemmatwo{}, we prove Proposition~\ref{it:at_eq}.
The proofs of all propositions translate well into Isabelle, so we refer to the original paper~\cite{new_fast_tableau} for the remaining proofs. 
\begin{proof}[Proof of Proposition~\ref{it:at_eq}]
  Assume that \lstinline!s $\eqS$ t! is in \lstinline!b!.
  If there exists a \lstinline!c $\in$ pwits b! where \lstinline!s = Var c! or \lstinline!t = Var c!, we arrive at a contradiction due to \reflemmatwo{}.
  Therefore, both \lstinline!s $\in$ subterms' b! and \lstinline!t $\in$ subterms' b! must hold.
  Now, assume for contradiction that \lstinline!realise s $\neq$ realise t!.
  Without loss of generality --- the other case is symmetric --- we obtain an \lstinline!e! such that
  \lstinline!e $\hfmem$ realise s! and \lstinline!e $\hfnotmem$ realise t!. 
  Considering that \lstinline!s $\in$ subterms' b! and the definition of \lstinline!realise!, we obtain a \lstinline!d! with \lstinline!e = realise d! and \lstinline!d $\rightarrow_\text{bgraph b}$ s!.
  This, in turn, yields that \lstinline!d $\inS$ s! must be in \lstinline!b!.
  Together with the assumption \lstinline!(s $\eqS$ t) $\in$ set b! and the saturation of \lstinline!b!, it follows that \lstinline!d $\inS$ t! must also be in \lstinline!b!.
  But then we have
  \lstinline!realise d $\hfmem$ realise t $\longleftrightarrow$ e $\hfmem$ realise t!
  using Proposition~\ref{it:at_mem}, which is a contradiction to the assumption \lstinline!e $\hfnotmem$ realise t!.
\end{proof}

\noindent We now lower the results on literals to set terms.
All of the proofs are straightforward so we refer the reader to the formalisation.
\begin{enumerate}[label=(\alph*)]
  \item\label{it:empty} It holds that \lstinline!realise $\emptyset$ = 0!.
  \item\label{it:op} Let \lstinline!$\star_\text{s}$ $\in$ {$\unionS$, $\diffS$, $\interS$}!. If the term \lstinline!s $\star_\text{s}$ t! occurs in \lstinline!subterms b!, then
    \begin{center}
    \lstinline!realise (s $\star_\text{s}$ t) = realise s $\star$ realise t!.
    \end{center}
  \item\label{it:single} If \lstinline!Single t $\in$ subterms b!, then
    \begin{center}
    \lstinline!realise (Single t) = HF {realise t}!.
    \end{center}
\end{enumerate}

The final step for obtaining a proper model is to connect the realisation function to the semantics as defined in \cref{sec:semantics}.
For set terms, we can use the Propositions~\ref{it:empty}--\ref{it:single} to prove the lemma below by induction on \lstinline!t!.
\begin{lstlisting}
lemma assumes t $\in$ subterms b
      shows &\Ist& ($\lambda$x. realise (Var x)) t = realise t
\end{lstlisting}
Lifting the above result to formulas yields the coherence of \lstinline!b!, as the original paper~\cite{new_fast_tableau} calls it.
The proof is a tedious but straightforward induction on the the size of the formulas.
\begin{lstlisting}[label={lst:coherence}]
lemma coherence:
  assumes $\phi$ $\in$ set b shows ($\lambda$x. realise (Var x)) $\models$ $\phi$
\end{lstlisting}
The coherence property finishes the proof of completeness of the calculus as it gives us a model for every formula in an open and saturated branch.

\section{Soundness of the Calculus\label[section]{sec:sound}}
A tableau calculus is sound if the corresponding formula is unsatisfiable for any closed tableau. 
We prove the following two properties to establish soundness:
\begin{enumerate}[label={(\arabic*)}]
  \item It is impossible to satisfy all formulas in a closed branch simultaneously. 
  \item The expansion rules maintain satisfiability.
\end{enumerate}

We formalise the first property in Isabelle below.

\begin{lstlisting}[belowskip=0pt,label={lst:bclosed_sound}]
lemma bclosed_sound:
  assumes bclosed b shows $\exists$$\phi$ $\in$ set b. M $\not\models$ $\phi$
\end{lstlisting}
\begin{proof}
  It is clear that, for any \lstinline!s!, neither does \lstinline!M! model \lstinline!s $\in$ $\emptyset$! nor \lstinline!s $\neqS$ s!.
  Furthermore, no model can satisfy both \lstinline!$\phi$! and \lstinline!&\fmNeg{$\phi$}&! at the same time.
  Lastly, a membership cycle is impossible since the membership relation of \lstinline!hf! is well-founded.
\end{proof}

\noindent We are left with showing that both linear and branching expansion rules preserve satisfiability.
As for the linear rules, a straightforward proof by case analysis on \lstinline!&\lexpands{b'}{b}&! suffices to obtain the lemma below.
\begin{lstlisting}[label={lst:lexpands_sound}]
lemma lexpands_sound:
  assumes &\lexpands{b'}{b}& and $\phi$ $\in$ set b' and $\bigwedge$$\psi$. $\psi$ $\in$ set b $\Longrightarrow$ M $\models$ $\psi$
  shows M $\models$ $\phi$
\end{lstlisting}
A similar argument would work for the branching rules if it were not for the last rule adding new variables.
Those variables need to be assigned specific values; hence, we modify the model as shown in the proof below.
\begin{lstlisting}[belowskip=0pt, label={lst:bexpands_sound}]
lemma bexpands_sound:
  assumes &\fexpands{bs'}{b}& and $\bigwedge$$\psi$. $\psi$ $\in$ set b $\Longrightarrow$ M $\models$ $\psi$
  shows $\exists$M'. $\exists$b' $\in$ bs'. $\forall$$\psi$ $\in$ set (b' @ b). M' $\models$ $\psi$
\end{lstlisting}
\begin{proof}
  We only consider the case where \lstinline!&\fexpands{bs'}{b}&! was proved by applying the last branching expansion rule to \lstinline!s $\neqS$ t! for some \lstinline!s! and \lstinline!t!.
  We have 
\begin{lstlisting}
  bs' = {[Var x $\inS$ s, Var x $\notinS$ t], [Var x $\inS$ t, Var x $\notinS$ s]}
\end{lstlisting}
  for some fresh variable \lstinline!x!.
  Since \lstinline!s $\neqS$ t! is in \lstinline!b!, we have that \lstinline!&\Ist& M s $\neq$ &\Ist& M t! because \lstinline!M! is a model.
  Without loss of generality, this inequality manifests itself through some \lstinline!y! with
  \lstinline!y $\hfmem$ &\Ist& M s! and \lstinline!y $\hfnotmem$ &\Ist& M t!.
  We update \lstinline!M! to map \lstinline!x! to \lstinline!y! to obtain the assignment \lstinline!M'!.
  Note that \lstinline!M'! is still a model for formulas in \lstinline!b! because \lstinline!x! is fresh with respect to \lstinline!b!.
  Furthermore, it is also a model for the first branch in \lstinline!bs'!, which finishes the proof.
\end{proof}

\section{Total Correctness of the Decision Procedure\label[section]{sec:correct}}
We first demonstrate the termination of the procedure for well-formed branches, i.e.\ every well-formed branch is in the domain of \lstinline!mlss_proc_branch!.
To this end, we derive an upper bound for the number of distinct formulas in a branch whose proof we omit here for brevity.
We should point out that this bound is not to be construed as the complexity of the procedure as it may create exponentially many branches in general.
\begin{lstlisting}[label={lst:card_branch}]
lemma card_wf_branch_ub:
  assumes wf_branch b
  shows |set b| $\leq$ 2$\:$*$\,$|subfms (last b)| + 16$\:$*$\,$|subterms (last b)|$^\text{4}$
\end{lstlisting}
Remember that \lstinline!mlss_proc_branch! only applies a linear expansion rule to a branch if the application results in new formulas.   
Moreover, the subsumption conditions of the branching expansion rules ensure that each of the newly created branches contain new formulas.
Ultimately, we conclude that the procedure must terminate for well-formed branches because the number of formulas increases in each step but is also bounded. 
\begin{lstlisting}
lemma assumes wf_branch b shows mlss_proc_branch_dom b
\end{lstlisting}

The above lemma allows us to utilise the computation induction rule of \lstinline!mlss_proc_branch! on well-formed branches, which we use to prove soundness and completeness.
As both proofs are essentially an application of soundness, respectively completeness, of the calculus, we refer the reader to the formalisation.
\begin{lstlisting}
lemma mlss_proc_branch_complete:
  fixes b :: 'a branch
  assumes wf_branch b and $\neg$ mlss_proc_branch b
  assumes infinite (UNIV :: 'a set)
  shows $\exists$M. M $\models$ last b

lemma mlss_proc_branch_sound:
  assumes wf_branch b and $\forall$$\psi$ $\in$ set b. M $\models$ $\psi$
  shows $\neg$ mlss_proc_branch b
\end{lstlisting}

To finish the proof of total correctness, note that every singleton branch is trivially well-formed; thus, termination, completeness, and soundness easily transfer to \lstinline!mlss_proc!.
\begin{lstlisting}
theorem mlss_proc_complete:
  fixes $\phi$ :: 'a pset_fm
  assumes $\neg$ mlss_proc $\phi$ and infinite (UNIV :: 'a set)
  shows $\exists$M. M $\models$ $\phi$

theorem mlss_proc_sound:
  assumes M $\models$ $\phi$ shows $\neg$ mlss_proc $\phi$
\end{lstlisting}

\section{Dealing with Urelements\label[section]{sec:urelements}}
In the introduction, we stated the goal of integrating \lstinline!mlss_proc! as a tactic into Isabelle.
For this to work, we must map every branch expansion rule to a corresponding theorem in Isabelle/HOL.
This is straightforward for all expansion rules except for the last branching expansion rule.
To illustrate, suppose that we are to disprove a statement of the form
\begin{lstlisting}
  s $\neq$ (t :: 'a) $\land$ s $\in$ (A :: 'a set) $\cup$ B $\land$ $\ldots$
\end{lstlisting}
in Isabelle/HOL.
By way of reification, we convert this to a formula of the shape
\begin{lstlisting}
  s' $\neqS$ t' $\fmAndSymbol$ s' $\inS$ A' $\unionS$ B' $\fmAndSymbol$ $\ldots$
\end{lstlisting}
in our set syntax for some \lstinline!s'!, \lstinline!t'!, \lstinline!A'!, and \lstinline!B'!.
When we apply the decision procedure to this formula, it might return a tableau proof that contains an application of the last branching rule to \lstinline!(s' $\neqS$ t') $\in$ set b!.
This results in two branches, one of which is \lstinline![Var x $\inS$ s', Var x $\notinS$ t'] @ b!;
however, there is no matching rule in Isabelle/HOL since \lstinline!s! and \lstinline!t! are not sets.

\begin{figure}[b]
  \centering
  \lstset{
    escapeinside={@}{@}
  }
  \begin{tabular}{c}
    \begin{tabular}{ccccc}
      &\qquad\quad&&\qquad\quad& \lstinline!$\Gamma$ $\vdash$ t : l! \\
      \cmidrule{1-1}\cmidrule{3-3}\cmidrule{5-5}
      \lstinline!$\Gamma$ $\vdash$ $\emptyset$ n : Suc n! & &
      \lstinline!$\Gamma$ $\vdash$ Var x : $\Gamma$ x! & &
      \lstinline!$\Gamma$ $\vdash$ Single t : Suc l! \\
    \end{tabular}
    \\[2em]
    \begin{tabular}{c}
      \lstinline!$\star_\text{s}$ $\in$ {$\unionS$,$\,\interS$,$\,\diffS$}! \quad
      \lstinline!$\Gamma$ $\vdash$ s : l! \quad
      \lstinline!$\Gamma$ $\vdash$ t : l! \quad 
      \lstinline!l $\neq$ 0! \\
      \midrule
      \lstinline!$\Gamma$ $\vdash$ s $\star_\text{s}$ t : l! \\
    \end{tabular}
    \\[2.5em]
    \begin{tabular}{ccc}
      \lstinline!$\Gamma$ $\vdash$ s : l! \quad \lstinline!$\Gamma$ $\vdash$ t : l! &\qquad\quad&
      \lstinline!$\Gamma$ $\vdash$ s : l! \quad \lstinline!$\Gamma$ $\vdash$ t : Suc l!
      \\
      \cmidrule{1-1}\cmidrule{3-3}
      \lstinline!$\Gamma$ $\vdash$ s $\eqS$ t! && \lstinline!$\Gamma$ $\vdash$ s $\inS$ t! \\
    \end{tabular}
  \end{tabular}
  \caption{The type system for set terms and atoms.\label[figure]{fig:types}}
\end{figure}

To deal with this problem, we formalise a lightweight type system as displayed in \cref{fig:types}. 
The type of a set term in this system is just a natural number which we call level.
Intuitively speaking, the level \texttt{l} means that the corresponding term \lstinline!t! in Isabelle/HOL has type
\[
  \texttt{'a}\ \underbrace{\texttt{set}\ \ldots\ \texttt{set}}_{\texttt{l}\ \text{times}}
\]
for some \lstinline!'a!.
Note that the constructor $\emptyset$ now receives an additional argument indicating the level of each instance of $\emptyset$.

Moreover, the typing judgement extends to set atoms by matching up the levels of its component set terms.

Ultimately, we define \lstinline!$\Gamma$ $\vdash$ $\phi$ $\equiv$ $\forall$a $\in$ atoms $\phi$. $\Gamma$ $\vdash$ a! in order to type formulas.

We can now define the urelements with respect to a formula.
An urelement is a set term whose corresponding type in Isabelle/HOL might not be a set.
\begin{lstlisting}
definition urelem :: 'a pset_fm $\Rightarrow$ 'a pset_term $\Rightarrow$ bool where
  urelem $\phi$ t $\equiv$ $\exists\Gamma$. $\Gamma$ $\vdash$ $\phi$ $\land$ $\Gamma$ $\vdash$ t : 0
\end{lstlisting}
Using this definition, we make two changes to the specification of the calculus: 
\begin{enumerate*}[label=(\arabic*)]
  \item First and foremost, we require that neither \lstinline!s! nor \lstinline!t! is an urelement in the precondition of the last branching expansion rule.
  \item As mentioned above, we add an argument to the $\emptyset$ constructor.
    This argument is only used for the typing judgement; it has no impact on the semantics.
\end{enumerate*}

Soundness, of course, is not affected by these changes but we have to make a few amendments to maintain completeness:
\begin{enumerate*}[label=(\arabic*)]
  \item The first equation of \lstinline!realise! now also must account for the urelements.
    In particular, it has to ensure that urelements receive pairwise different values unless they are related through equality atoms. 
    This does not affect pure witnesses since they can not be related through equality atoms due to \reflemmatwo{}.
  \item We must adjust the completeness proof in those places where it directly refers to the definition of \lstinline!realise! to account for the case where a given term is an urelement.
  \item The completeness theorem receives the additional assumption that \lstinline!$\Gamma$ $\vdash$ $\phi$! holds for the initial formula $\phi$. 
  \item For the completeness proof, we must show that the typing judgement is invariant under branch expansion.
\end{enumerate*}

The modifications above ensure that the proof can be replayed through Isabelle/HOL.
To actually use the calculus, we must determine the urelements of the initial formula $\phi$, though.
In other words, we have to implement an inference algorithm for our lightweight type system.
The algorithm is, in essence, a simplified version of Hindley-Milner type inference so it has the same two phases:
it generates constraints using syntax directed rules and then passes them to a constraint solver.

Since we are only interested in the level of a term, we can encode all constraints into the theory of $0$, the successor function \lstinline!S!, and equality (but no disequality).
Note that constraints of the form \lstinline!l $\neq$ 0! can be replaced by \lstinline!l $=$ S i! with \lstinline!i! being a fresh variable.
A solver for this theory is straightforward to implement and verify;
nevertheless, we have to be careful that it computes the minimum assignment $\Gamma$ from variables to levels that fulfils the constraints. 
This guarantees that a set term \lstinline!t! is not an urelement if, and only if, \lstinline!$\Gamma$ t $>$ 0!.
Conversely, all terms \lstinline!s! with \lstinline!$\Gamma$ s $=$ 0! are urelements.

\section{Conclusion and Future Work}
We developed a formalisation of a tableau calculus for a quantifier-free fragment of set theory called \MLSS{} based on a paper by \citeauthor{new_fast_tableau}~\cite{new_fast_tableau}. 
The formalisation includes an abstract description of a decision procedure that builds on the calculus.
To make the decision procedure compatible with Isabelle/HOL, we extended the calculus with a lightweight type system while maintaining completeness.
We also refined the abstract specification to an executable specification from which code can be generated.

In future work, we plan to implement an efficient executable specification in the style of a worklist algorithm. 
This specification should also generate certificates that can be replayed through Isabelle's inference kernel to facilitate the integration of the procedure into Isabelle.


\subsubsection{Acknowledgements}
The author thanks Kevin Kappelmann and Tobias Nipkow for their comments on a draft version of this paper and the anonymous referees for their thorough reviews.

\bibliography{sources}


\end{document}